\documentclass[final]{IEEEtran}

\usepackage{amsthm,amssymb,graphicx,multirow,amsmath,color,amsfonts}%,ulem}
\usepackage[update,prepend]{epstopdf}
\usepackage[noadjust]{cite}
\usepackage[latin1]{inputenc}
\usepackage{tikz}
\usetikzlibrary{arrows,calc}		% Optional ticks libraries
\usepackage{bbm} % for \mathbbm{1}
\usepackage{pdfpages}
\usepackage{tabulary}
\usepackage{multirow}
\usepackage{comment}
\def\chb#1{{\color{black} #1}}
\def\chm#1{{\color{black} #1}}
\def\chg#1{{\color{black} #1}}

\def\nby{{\mathbf{y}}}

\def\nb0{{\mathbf{0}}}
\def\nb1{{\mathbf{1}}}

\def\ncalB{{\mathcal{B}}}

\def\ncalI{{\mathcal{I}}}

\def\ncalL{{\mathcal{L}}}

\def\nrmd{{\rm d}}

\newtheorem{lemma}{Lemma}
\newtheorem{thm}{Theorem}

\newtheorem{cor}{Corollary}

\newtheorem{remark}{Remark}

\def\figref#1{Fig.\,\ref{#1}}%
\def\E{\mathbb{E}}

\def\P{\mathbb{P}}

\def\T{\beta}							% Threshold = \beta_i
\def\sir{\mathtt{SIR}}

\allowdisplaybreaks % Allows breaking of eqnarray over multiple pages (avoids unnecessary blanks in the document before eqnarray)

\begin{document}
\graphicspath{{./Figures/}}
\title{
 Optimal   Geographic Caching in Finite Wireless Networks}
\author{Mehrnaz Afshang and Harpreet S. Dhillon
\thanks{The authors are with Wireless@VT, Department of ECE, Virgina Tech, Blacksburg, VA, USA. Email: \{mehrnaz, hdhillon\}@vt.edu. The support of the US NSF (Grant CCF-1464293) is gratefully acknowledged.} }

\maketitle

\begin{abstract}
\chb{Cache-enabled device-to-device (D2D) networks turn memory of the devices at the network edge, such as smart phones and tablets, into {\em bandwidth} by enabling \chm{asynchronous content sharing directly} between \chm{proximate} devices.} Limited storage capacity of the mobile devices necessitates the determination of optimal set of   contents to be cached on each device. In order to study the problem of optimal cache placement, we model the locations of devices in a finite region (e.g., coffee shop, sports bar, library) as a  uniform binomial point process (BPP). For this setup, we first develop a generic framework to analyze the coverage probability of the target receiver (target-Rx) when the requested content is available at the $k^{th}$ closest device to it. Using \chm{this} coverage probability result, we evaluate  optimal caching probability of the popular content to maximize the total hit probability. Our analysis concretely demonstrates that optimal caching probability  \chg{strongly} depends on the number of simultaneously active devices in the network.

\end{abstract}

\begin{IEEEkeywords}
Optimal cache placement, $k$-coverage analysis, D2D networks, and binomial point process.
\end{IEEEkeywords}

\section{Introduction} \label{sec:intro}
\chb{Video-driven applications, such as on-demand video streaming and mobile television, are primary} drivers of the explosion in mobile data traffic~\cite{cisco2014global}.  Empirical studies  of the distribution of  video requests   reveals  a high degree of  spatiotemporal correlation in the content demanded, which means  that a few popular \chb{videos}  are requested by multiple users at different times~\cite{cha2007tube}. However, the current cellular \chg{network} do  not  exploit  this  spatiotemporal correlation exhibited in the content demand, which can  be effectively used  as an active part of the future networks  design. In particular, popular contents  can be locally cached at the edge of the network such as \chb{in small cell base stations (BSs)} or mobile devices and then delivered  over a direct link to the user who requests  them~\cite{golrezaei2013femtocaching, bastug2014living,KrishAfshDhillon2016}. 
\chb{However, limited storage capacity of the edge devices necessitates the need for developing optimal caching strategies. While this problem has been addressed for infinite networks in~\cite{BlaszczyszynG14,ShDhiC2015a,yang2016analysis, bacstu?2015cache, malak2014optimal}, more relevant D2D use case of finite networks has not yet been addressed in the literature.}
The modeling and performance analysis of optimal geographical caching in finite wireless networks is  \chb{therefore the} main focus of this paper.

 {\em Related work.} Modeling and analysis of the cache-enabled heterogeneous networks (HetNets) including \chb{small cell BSs} and mobile devices have taken two main directions in the literature. The first line of work focuses on the \chb{fundamental throughput scaling results} by assuming simple protocol channel model, where communication between two nodes is successful if they are within the fixed collaboration distance \cite{shanmugam2013femtocaching,6787081}.   The second line of work which is also  relevant to the current work considers a more realistic model for the underlying physical layer, where the successful communication depends on physical quantities such as \chb{the} path-loss, fading, and interference power. More specifically, these works focus on optimal cache-placement strategies in D2D or ultra-dense HetNets by using the tools from stochastic geometry \cite{BlaszczyszynG14,ShDhiC2015a,yang2016analysis, bacstu?2015cache, malak2014optimal}. The most \chb{popular technical} approach amongst these works is modeling the locations of nodes (devices or small cell BSs) as \chb{an infinite} homogenous Poisson Point Process (PPP) and \chm{assume that each device connects to its closest possible serving node}. The main reason behind the popularity of this approach is the simplicity involved in the mathematical analysis as demonstrated in \cite{AndrewsTractable}. 
 This setup  \chb{\em albeit} relevant \chg{for} the traditional cellular architecture, is not adequate in D2D and ultra-dense cache-enabled HetNets where the content of interest may not necessarily be available at the closest node (device or small cell BS) to the target-Rx. \chm{Relaxing this constraint, \cite{BlaszczyszynG14} \chg{defined} an optimization problem that determines the optimal caching strategy using the $k$-coverage results for an infinite PPP derived in~\cite{H.P.Keelerkcoverage}.} \chb{In continuation to \cite{BlaszczyszynG14}, we determine the optimal geographic caching strategy for finite wireless networks in this paper. To the best of our understanding, this is the first work on optimal caching in finite networks.} More details are provided next.

{\em Contributions and outcomes.} We model the locations of devices in a  finite wireless network as   a uniform-BPP. For this setup,  we develop a new set of  tools to characterize the  ``exact'' expression of the coverage probability where the serving device is \chb{the} $k^{th}$ closest device to the target-Rx out of \chb{the} $N^{\tt t}$ transmitting devices.  To perform this analysis, we prove that the distance from interfering devices conditioned on the location of \chb{the} serving device are independently and identically distributed (i.i.d.). Using this i.i.d. property, we derive a closed-form expression for the Laplace transform of interference distribution that is the main component of the coverage probability analysis. \chm{This coverage probability result for finite networks is analogous to the $k$-coverage result of~\cite{H.P.Keelerkcoverage} and can be easily incorporated into the optimization formulation of \cite{BlaszczyszynG14} in order to characterize throughput and determine the optimal caching strategy that maximizes total hit probability in finite networks.}
\chm{Our analysis demonstrates that the increasing number of active devices} has a conflicting effect on the maximum hit probability and the throughput: maximum hit probability decreases and throughput increases. This shows that more and more  devices can be simultaneously activated as long as the hit probability remains acceptable.

%%%%%%%%%%%%%%%%%%%%%%
\section{System Model} \label{sec:SysMod}
\subsubsection*{System setup and key assumption}
We consider a  cache-enabled D2D network comprising $N^{\tt t}+1$ devices that \chb{include} target-Rx and $N^{\tt t}$ transmitting devices.  Modeling the locations of transmitting devices as a uniform-BPP $\{{\bf y}_i\} \equiv \Phi_{\tt t}$, we assume that $N^{\tt t}$ transmitting devices are independently and uniformly distributed  in  the ball of radius $r_{\rm d}$ centered at origin, which is denoted by ${\bf b}(0, r_{\rm d})$.  We further assume that $N^{\rm a}$ out of $N^{\tt t}$  transmitting (serving and interfering) devices simultaneously reuse the same resource block. The locations of simultaneously active devices  is denoted by $\Phi_{\rm a}$. For this setup, we perform analysis on a  target-Rx located at  origin.
 Note that while our analysis is,
in principle, extensible to any arbitrarily located target-Rx,  we limit it to this case  due space  constrains. The analysis corresponding to the arbitrarily located target-Rx will be provided in the extended version of this paper. 

\subsubsection*{Channel model} \chm{In order to pose an optimization problem similar to \cite{BlaszczyszynG14} for finite networks, we are interested in the case where the content of interest for the target-Rx is available at \chg{its} $k^{th}$ closest device out of $N^{\tt t}$ transmitting devices, \chg{ where $1\leq k\leq N^{\tt t}$.}} Fixing the location of  this serving device which is denoted by ${\bf y}_k$, we assume that \chm{the} interfering devices,  located at ${\bf y}_i \in \Phi_{\rm a} \setminus {\bf y}_k \subset \Phi_{\tt t}$, are chosen uniformly at random from \chb{the} set of transmitting devices.  To keep the notation simple, we assume that the \chg{thermal} noise is negligible  compared to the interference and is hence ignored. So, the signal to interference ratio ($\sir$) is:

 \begin{equation}\label{eq: SIR}
 \sir=\frac{   h_k \|{\bf y}_k\|^{-\alpha}}{\sum_{{\bf y}_i\in \Phi_{\rm a}\setminus {\bf y}_k}  h_i\|\nby_i\|^{-\alpha}},
  \end{equation}
  where $h_i\sim \exp(1)$ and $\|.\|^{-\alpha}$ model Rayleigh fading and
power law path-loss \chm{with exponent $\alpha>2$}, respectively.
\begin{remark}[Scale invariance of BPP network]   The $\sir$ \chb{as defined in \eqref{eq: SIR}} is independent of the network radius $r_{\rm d}$ because \chb{both serving and interfering
devices} are chosen from the same \chb{point} process. This is mainly because by changing $r_{\rm d}$, all distances scale with the same parameter.   So, without loss of  generality, we normalize \chb{$r_{\rm d}$ to $1$}.
\end{remark}
  \subsubsection*{ Content popularity}
    We consider a finite  library of popular contents ${\cal C}=\{{ c}_1,c_2,..., c_{\cal J}\}$ with size ${\cal J}$, where $c_j$ denotes the \chg{$j^{th}$} most popular content. For simplicity of exposition, we  assume that all the content has the same size, \chb{which} is normalized to one. We further assume that the transmitting devices are equipped with cache storage of size $m_{\rm c}$ and hence each device can  \chb{store at most} $m_{\rm c}$ popular contents. Now, denoting the specific set of contents at  a generic  transmitting device \chb{by} $\Omega$, the caching probability  is:
    $b_{j}=\P(c_j\in \Omega), $
  where $b_{j}$ denotes the probability that  the content $c_j$ is stored at a given transmitting device.
To model the content popularity in this system, we use  Zipf's distribution due to its practical relevance \cite{cha2007tube}. Hence,  the request probability for file $c_{j}$ is:
\begin{align}\label{eq: zipf}
{\tt P}_{R_j}=\frac{j^{-\gamma}}{\sum_{i=1}^{\cal J} i^{-\gamma}}; \quad 1 \le j \le {\cal J},
\end{align}
where $\gamma$ represents the parameter of Zipf's distribution.   
  \begin{figure}[t!]
\centering{
        \includegraphics[width=.43\linewidth]{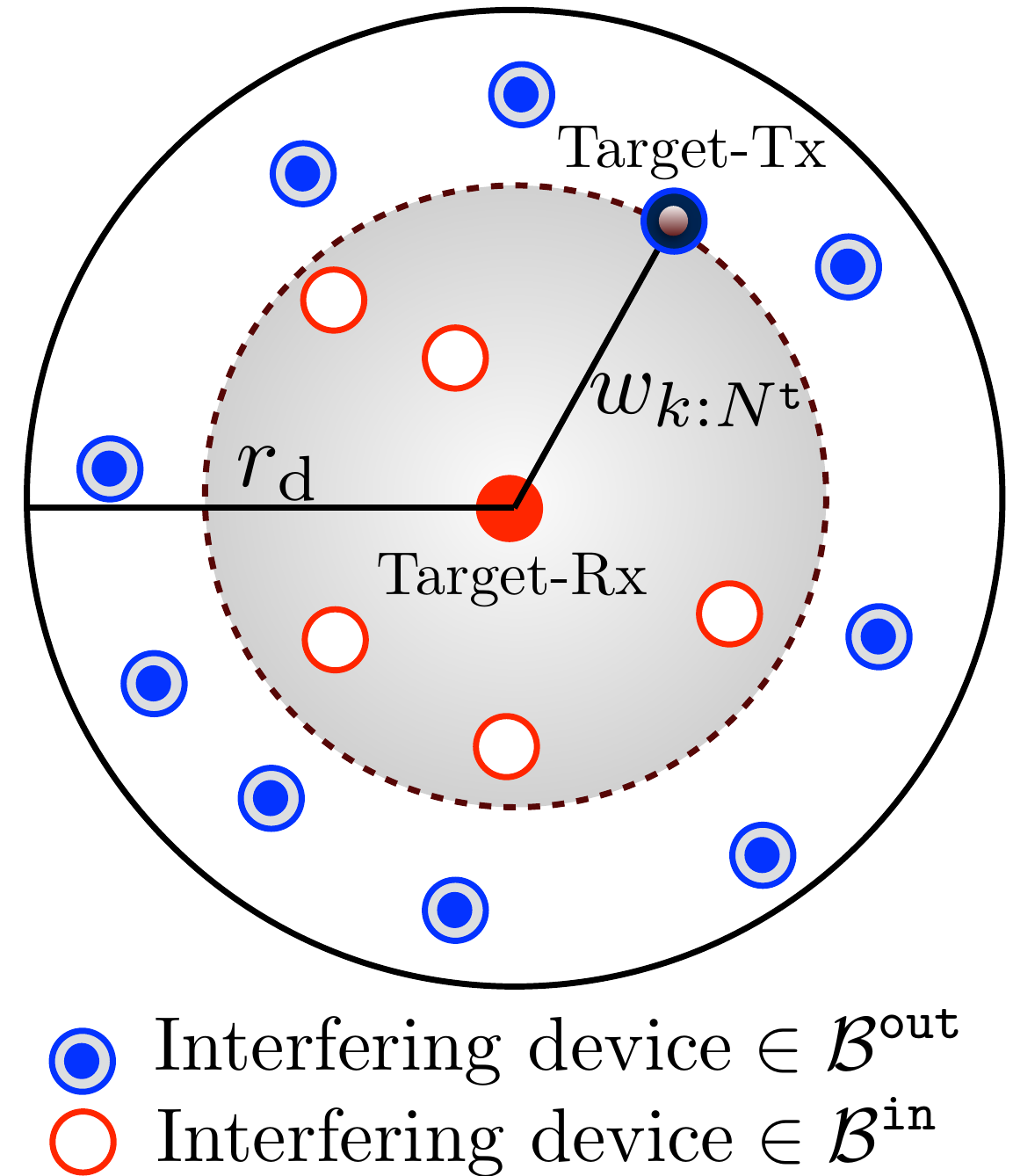}
              \caption{Illustration of the system model.}
                \label{Fig:sys_K}
                }
\end{figure}
 \section{Optimal content  placement}
This is the main technical section of this paper, where we characterize  network performance in terms of total hit probability, coverage probability, and network throughput. Before going into the detailed analyses of these metrics, we   first characterize the distribution of
the distances from the target-Rx to the transmitting  devices in the next subsection.
\subsection{Relevant Distance Distributions in a BPP}
\label{subsec: On the Distribution of Distance in BPP}
In our setup, the target-Rx is assumed to be located at the origin \chb{while the} transmitting devices are uniformly and independently distributed  in ${\bf b}(0,r_d)$. Therefore, it is easy to infer that the  sequence of distances from transmitting devices to the target-Rx which is denoted by $\{W_i=\|{\bf y}_i\|\}$ \chb{contains i.i.d. elements} \chg{with PDF and CDF given by~\cite{srinivasa2010distance}:}
\begin{align}\label{Eq:PDF unf}
\text{PDF}: \quad f_{W_i}(w_i)=\frac{2 w_i}{r_{\rm d}^2}; \quad 0 \le w_i \le r_{\rm d},\\ \label{Eq: CDF unif}
\text{CDF}: \quad F_{W_i}(w_i)=\frac{w_i^2}{r_{\rm d}^2}; \quad 0 \le w_i \le r_{\rm d}.
\end{align}
Recall that the serving device is \chb{the} $k^{th}$ \chg{closest}  transmitting device to the target-Rx. In order to characterize the serving distance distribution, we  need to ``order" the distances from
transmitting devices to the target-Rx.  We define an ordered set $\{w_{i:N^{\tt t}}\}_{i=1:N^{\tt t}}$ by sorting $w_i$-s in ascending order such that  $w_{1:N^{\tt t}}<w_{2:N^{\tt t}}<...< w_{N^{\tt t}:N^{\tt t}}.$
Now, using the  i.i.d. property of the sequence $\{W_i\}$, the PDF of \chb{the} serving distance  $R=W_{k: N^{\tt t}}$ is: 
\begin{multline}
f_R^{(k)}(r)=\frac{N^{\rm t}!}{(k-1)!(N^{\rm t}-k)!}
 {F_{W_i}(r)}^{k-1} \\ \times f_{W_i}(r)(1-F_{W_i}(r))^{N^{\rm t}-k},
  \label{eq: serving k policy part}
  \end{multline}
 where, $f^{(k)}_R(r)$ can be \chb{obtained simply} from the PDF of the $k$ order statistics of the sequence $\{W_i\}$ \cite{ahsanullah2005order}. It is worth highlighting that  the interfering devices can lie at  any place \chm{except} the location of the serving device. This means that the $k^{th}$ closest transmitting device to the target receiver is  explicitly removed from \chb{the} interference field. \chb{In order to  incorporate this in the analysis, we partition} the set of distances from transmitting devices to the target-Rx into three subsets ${\cal B}=\{\ncalB^{\tt in},w_{k:N^{\tt t}}, \ncalB^{\tt out} \}$  such  that $\ncalB^{\tt in}$ and $\ncalB^{\tt out}$ represent the set of potential  interfering devices closer and farther to the target-Rx respectively, compared to the serving device. This setup is illustrated  in \figref{Fig:sys_K}.
  The  following  \chb{Lemma deals} with conditional i.i.d. property of $U_{\tt in} \in \ncalB^{\tt in}$ and $U_{\tt out}\in \ncalB^{\tt out}$, and \chm{provides} their density functions. The proof is provided in Appendix \ref{App: i.i.d. property of u_in and u_out}.
  \begin{lemma}\label{lem: density function of interferer distance k}
 The \chb{sequences} of random variables $U_{\tt in} \in \ncalB^{\tt in}$   and $U_{\tt out} \in \ncalB^{\tt out}$ conditioned on  $r=w_{k:N^{\tt t}}$ are  independent. Moreover, 
  
  i)  the \chb{elements in}  sequence $U_{\tt in} \in \ncalB^{\tt in}$  conditioned on  $r=w_{k:N^{\tt t}}$ are i.i.d., where the PDF of each element  is:
  \begin{align} \label{eq: pdf uin}
  f_{U_{\tt in}}(u_{\tt in}| r)=\left\{
 \begin{array}{cc}
 \frac{f_{W_i}(u_{\tt in})}{F_{W_i}(r)}, &u_{\rm in}<r  \\
 0, & u_{\tt in}\geq r
 \end{array}\right.,
  \end{align} 
  
  ii) the   \chb{elements in}  sequence  $U_{\tt out} \in \ncalB^{\tt out}$  conditioned on  $r=w_{k:N^{\tt t}}$ are i.i.d., where the PDF of each element is:
\begin{align}
  f_{U_{\rm out}}(u_{\tt out}| r)=\left\{
 \begin{array}{cc}
 \frac{f_{W_i}(u_{\tt out})}{1-F_{W_i}(r)}, & u_{\tt out}> r\\
 0, & u_{\rm out}\le r
 \end{array}\right.,
  \end{align}
 where $f_{W_i}(.)$ and $F_{W_i}(.)$  are given by  \eqref{Eq:PDF unf} and \eqref{Eq: CDF unif}, respectively.
  \end{lemma}
%  \begin{IEEEproof}
%  See Appendix \ref{App: i.i.d. property of u_in and u_out}.
%  \end{IEEEproof}
In \cite{MehrnazD2D1,afshang2015fundamentals}, we proved a similar i.i.d. property for  the distribution of distances in Thomas cluster process. The  conditional i.i.d. property of $U_{\tt in} \in \ncalB^{\tt in}$   and $U_{\tt out} \in \ncalB^{\tt out}$ is the key enabler \chb{for the analysis of the Laplace transform of interference, which is discussed in the next subsection.}
\subsection{Laplace Transform of Interference Distribution}
  As will be evident in the sequel,  the characterization of  Laplace transform of interference distribution is the  key intermediate step in the coverage probability analysis. The   {``exact''}  closed-form expression of the Laplace transform of interference  where serving device is \chb{the} $k^{th}$ closest transmitting device to the target-Rx is given in the next \chb{Lemma}. The proof is provided in Appendix \ref{App : Laplace k-closet}.
\begin{lemma}\label{lemma: Laplace under k-closest}
The Laplace transform of interference distribution conditioned on the serving distance $r=w_{k:N^{\tt t}}$ is $\ncalL_{\ncalI}^{(k)}(s|r)=$
\begin{align}\notag
&\sum_{\ell=0}^{n^{\rm a}_{m}}  \xi(p,n^{\rm a}_{m}) \left(\frac{{\cal C}(\alpha,s, r)}{r^2}\right)^{\ell} \left(\frac{{\cal C}(\alpha,s, r_{\rm d})-{\cal C}(\alpha,s, r)}{r_{\rm d}^2-r^2}\right)^{N^{\rm a}-\ell-1}\\\label{eq: Laplace k closest}
&\text{with} \quad {{\cal C}(\alpha, s, x)}=x^2-x^2 \:{}_2F_1(1, \frac{2}{\alpha}, 1+\frac{2}{\alpha},{-x^{\alpha}/ s})),
\end{align}
where $\xi(p,n^{\rm a}_{m})=\frac{p^\ell (1-p)^{N^{\rm a}-\ell-1} \binom{N^{\rm a}-1}{\ell}} {\sum_{\ell=0}^{N^{\rm a}_{m}}p^\ell (1-p)^{N^{\rm a}-\ell-1} \binom{N^{\rm a}-1}{\ell}}$, $p=\frac{k-1}{N^{\tt t}-1}$, $n^{\rm a}_{m}=\min(k-1,N^{\rm a}-1)$, and ${}_2F_1(a,b;c;z) = 1+\sum_{k=1}^\infty \frac{z^k}{k!}\prod_{l=0}^{k-1} \frac{(a+l)(b+l)}{c+l}$.
\end{lemma}
%\begin{IEEEproof}
%See Appendix \ref{App : Laplace k-closet}
%\end{IEEEproof}
       \begin{figure}[t!]
\centering{
        \includegraphics[width=.68\linewidth]{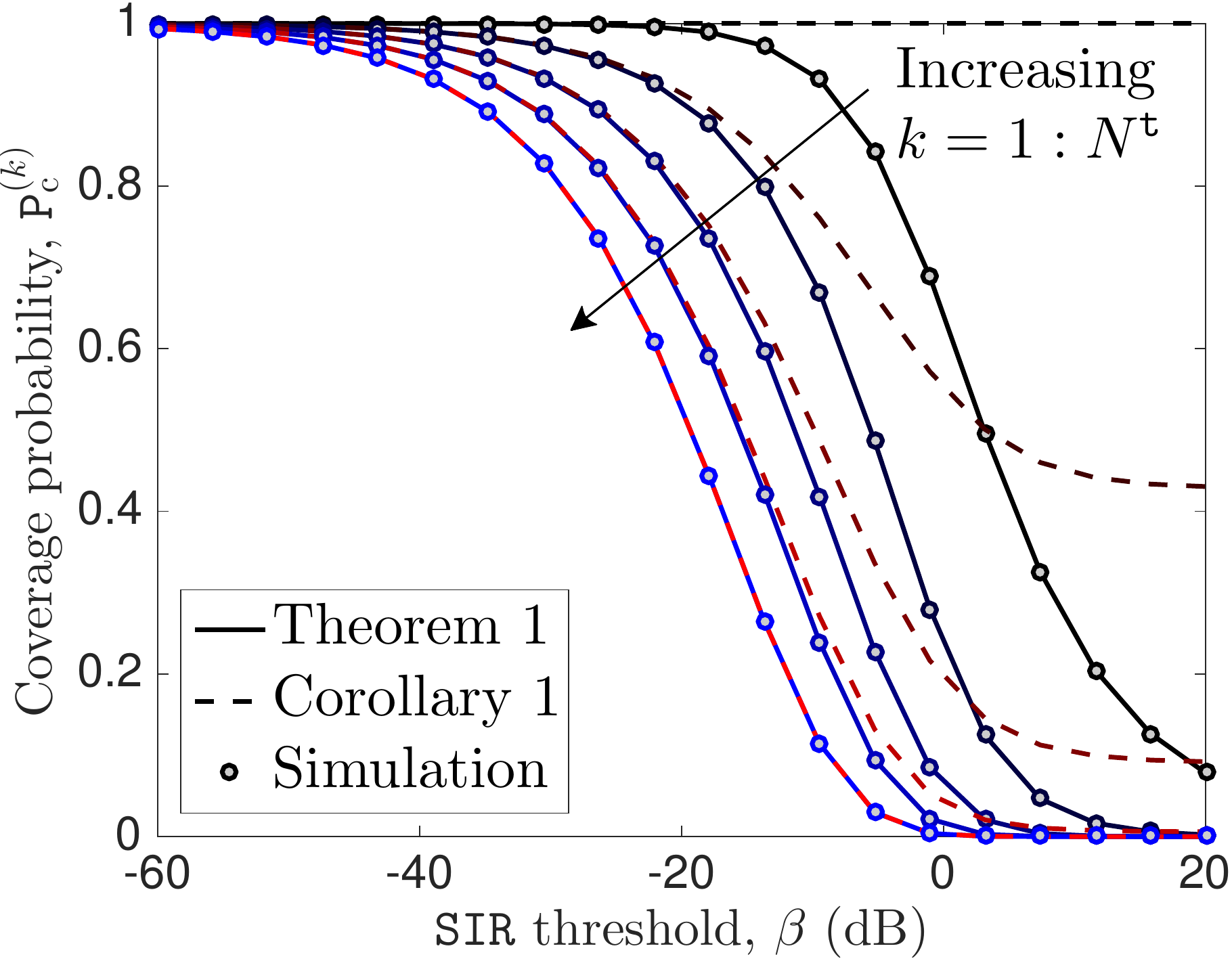}
              \caption{Coverage probability as a function of $\sir$ threshold ($\alpha=4$,  and  $N^{\rm a}=N^{\tt t}=5$). }
                \label{Fig: PC validation}
                }
\end{figure}
  \subsection{Coverage Probability }
  Coverage probability is defined as  the probability that  $\sir$ exceeds   predefined threshold needed to establish  a successful  connection. 
 % This can be mathematically expressed as: ${\tt P}_{\rm c}=\P(\sir>\beta)$.  
  The coverage probability of the target-Rx \chm{in a finite network} is derived in the next Theorem.
\begin{thm} Using  the expression of Laplace transform of interference  distribution given by  \eqref{eq: Laplace k closest}, the  coverage probability of the target-Rx  is:
\begin{align}
&{\tt P}_c^{(k)}=\int_0^{r_{\rm d}} \ncalL_{\ncalI}^{(k)}(\beta r^{\alpha}|r)f_R^{(k)}(r) {\rm d} r,   \label{eq: coverage k-closest central}
\end{align}
where $f_R^{(k)}(r)$ given by \eqref{eq: serving k policy part}, and $\T$ is target $\sir$ threshold. 
\label{thm: coverage k-closest}
\end{thm}
\begin{IEEEproof}Using  the definition of coverage probability, i.e., $\P(\sir\ge\T)=\P\big(h_k\ge \T r^\alpha{ \sum_{{\bf y}_i \in \Phi_{\rm a}\setminus {{\bf y}_{k}}} h_i\|\nby_i\|^{-\alpha}}\big)$
\begin{align*}
%pc=&\P(h_\ell\ge \T r^\alpha \sum_{i=1,i\neq \ell}^{N^{\rm t}}  h_i\|\nbx_0+\nby_i\|^{-\alpha})\\
\stackrel{(a)}=&\E_R\Big[\exp\Big(-\T r^{\alpha} {\sum_{{\bf y}_i \in \Phi_{\rm a} \setminus {\bf y}_k}  h_i\|\nby_i\|^{-\alpha}} \Big)\big |R\Big]
\end{align*}
where $(a)$ follows from $h_0\sim \exp(1)$. The \chg{final} result \chb{follows from} the definition of Laplace transform, followed by de-conditioning over serving distance $R$.
\end{IEEEproof}
  Using the result of Theorem \ref{thm: coverage k-closest}, we now derive a closed-form upper bound on the coverage probability in the next Corollary. The result can be readily derived by ignoring the contribution of interference from interfering devices farther than \chm{the} serving device to the target-Rx, and substituting $s=\beta r^{\alpha}$ in the expression of Laplace transform interference distribution given by  \eqref{eq: Laplace k closest}. %The tightness of the upper bound will be studied in the result and discussion section.
  \begin{cor}The coverage probability of a target-Rx is upper bounded by
  \begin{align}
  &{\tt P}_c^{(k)}\le \sum_{\ell=0}^{n^{\rm a}_{m}}  \xi(p,n^{\rm a}_{m})  \left((1- {}_2F_1(1, \frac{2}{\alpha}, 1+\frac{2}{\alpha},{-\frac{1}{ \beta}}))\right)^{\ell}\\\notag
&  \text{which for $N^{\rm a}=N^{\tt t}$ and $\alpha=4$  simplifies to}\\
%&{{\tt P}_{\rm c}}\le \left((1- {}_2F_1(1, \frac{2}{\alpha}, 1+\frac{2}{\alpha},{-1/ \beta}))\right)^{k-1}.
&{{\tt P}_{\rm c}}\le \left(1-\sqrt{\beta}\arctan\left(1/\sqrt{\beta}\right)\right)^{k-1}.
  \end{align}
  \label{for: upper bound}
    \end{cor}
    \begin{remark}From Corollary \ref{for: upper bound}, it is clear that the coverage probability has a power-law behavior with the exponent of $k-1$ when all transmitting nodes are simultaneously active.
%Furthermore, the bound gets tight by increasing $k$.
    \end{remark}
  We plot  the coverage probability for \chb{different values of} $k$ in \figref{Fig: PC validation}. It can be seen that the ``exact'' expression of the coverage probability given by Theorem \ref{thm: coverage k-closest}  and the simulation result match perfectly which confirms the accuracy of the analysis. \chg{As expected,} the closed-form expression given by Corollary \ref{for: upper bound} gets tight with increasing $k$ \chm{while being {\em exact}} for $k=N^{\tt t}$.  The upper bound, however, gets loose with increasing  target $\sir$ threshold.  This is mainly because the impact of interfering devices in ${\cal B}^{\tt out}$ on the coverage probability is more prominent for large $\sir$ threshold.
 \subsection{Total Hit Probability and Throughput} 
\begin{figure}[t!]
\centering{
        \includegraphics[width=.68\linewidth]{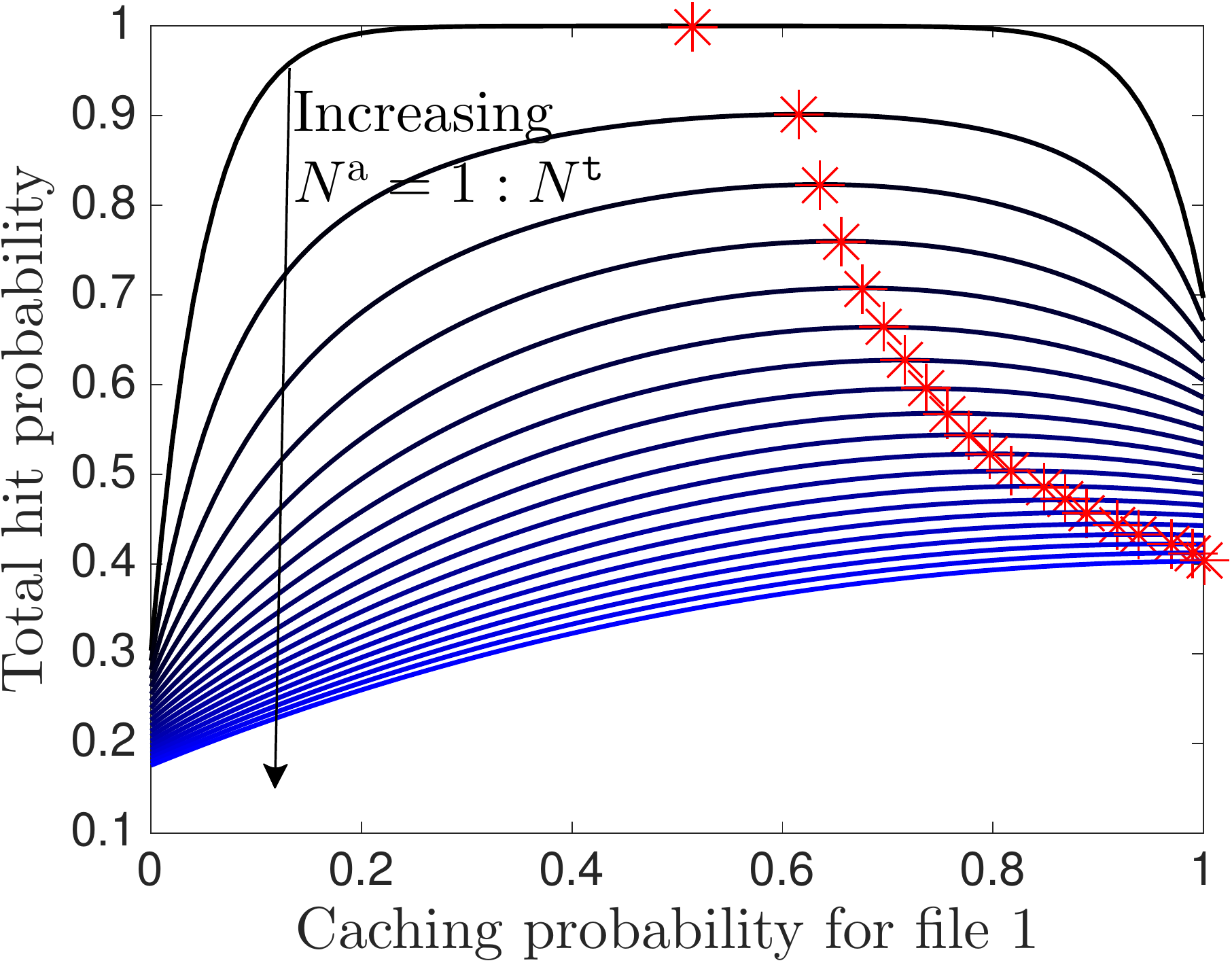}
              \caption{Total hit probability as  a function of caching probability for file 1, where asterisk shows optimal hit probability (${\cal J}=2$,  $m_{\rm c}=1$,  $\alpha=4$, $\beta=0$ dB, $\gamma=1.2$, and $N^{\tt t}=20$) }
                \label{Fig:optimal hit prob}
                }
\end{figure}

\begin{figure}[t!]
\centering{
        \includegraphics[width=.68\linewidth]{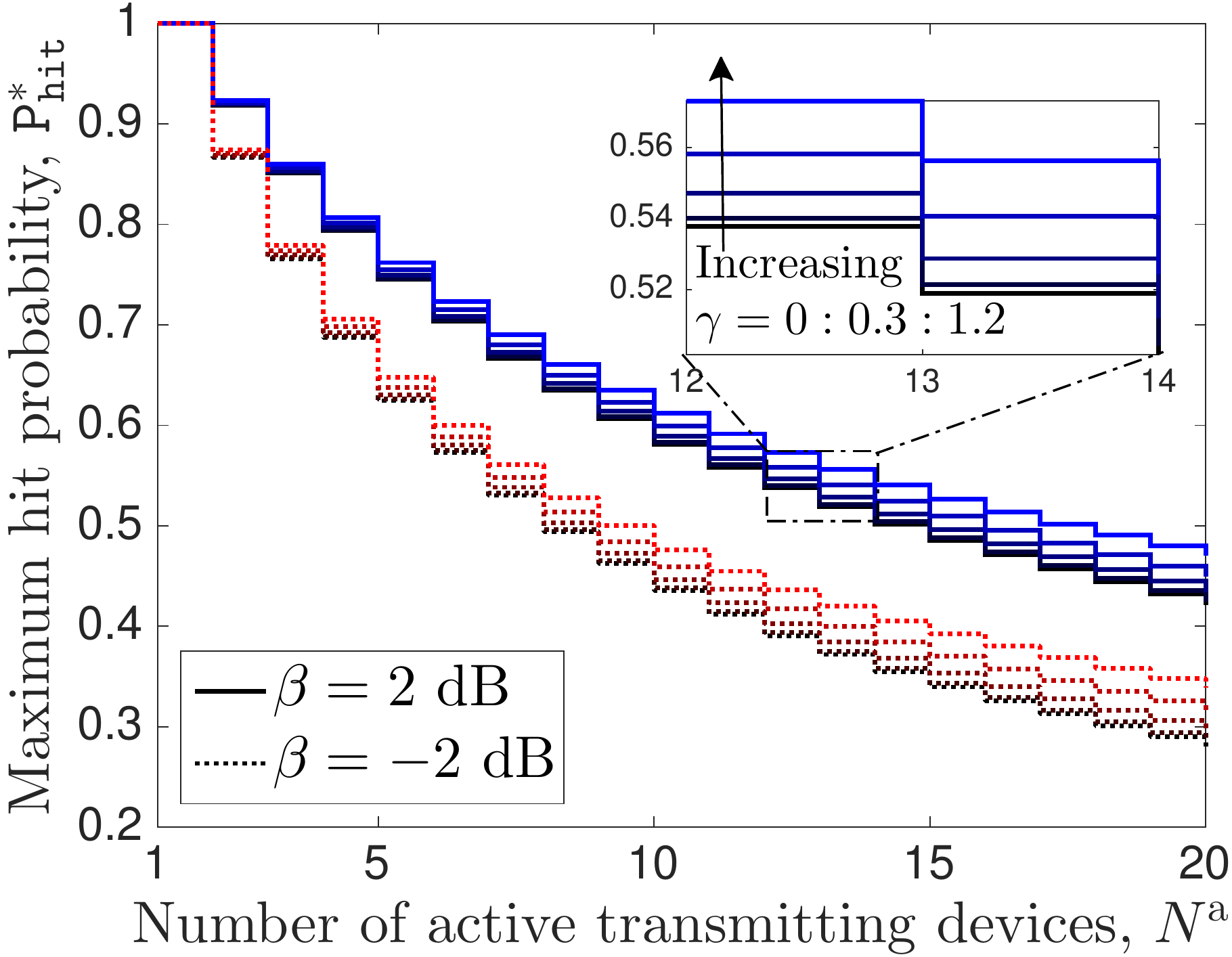}
              \caption{Maximum hit probability as a function of number of active devices (${\cal J}=2$, $m_{\rm c}=1$, $\alpha=4$, and $N^{\tt t}=20$)}
                \label{Fig:optimal hit prob vs number of active}
                }
\end{figure}

\begin{figure}[t!]
\centering{
        \includegraphics[width=.68\linewidth]{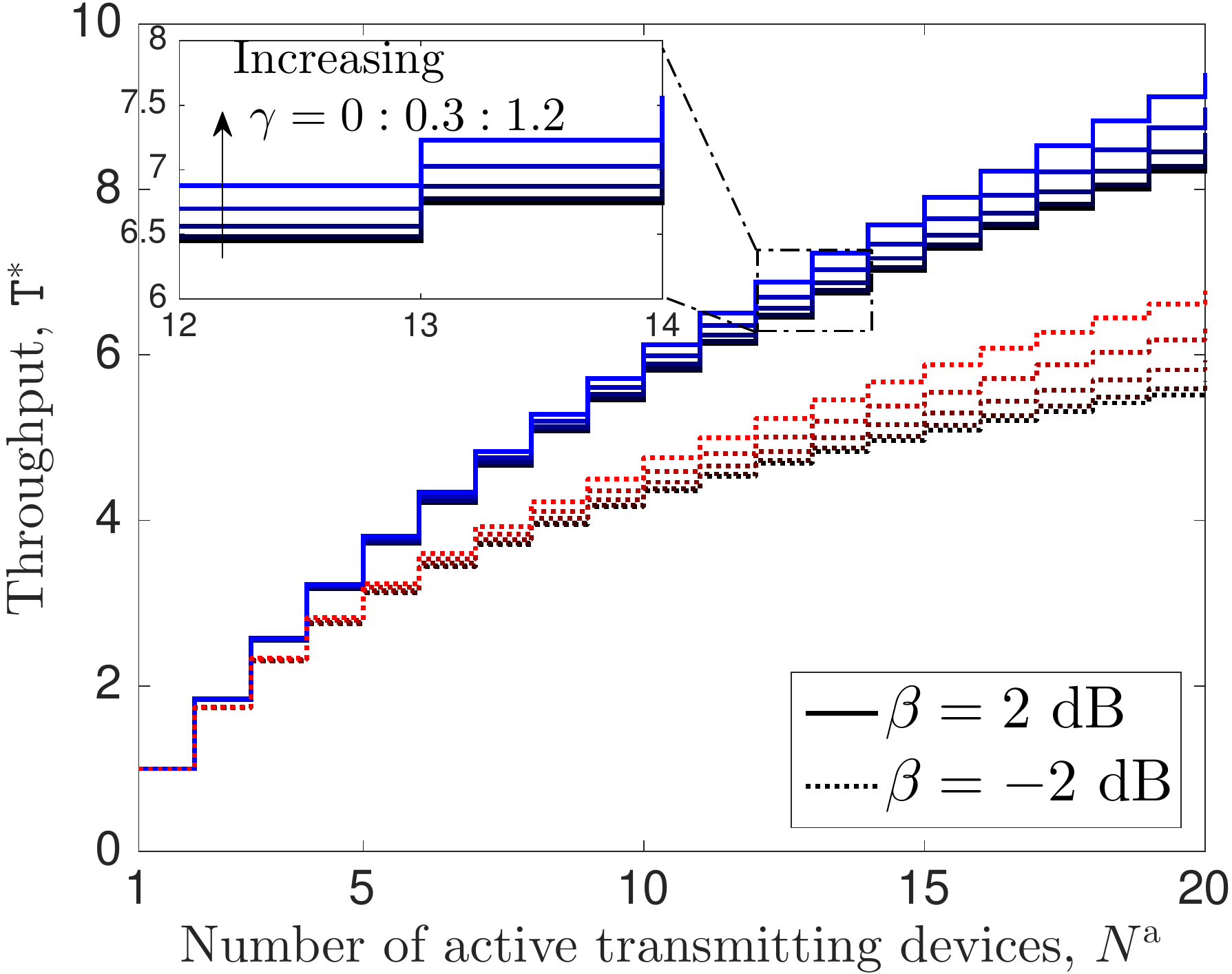}
              \caption{ Throughput as a function of number of active devices (${\cal J}=2$, $m_{\rm c}=1$, $\alpha=4$, and $N^{\tt t}=20$) }
                \label{Fig:optimal ASE vs number of active}
                }
\end{figure}

 \chm{As defined in [6], the {\em total hit probability is}} the probability that the target-Rx finds its requested content in one of the devices that cover it, which in turn depends on: i) request probability, \chm{which} is assumed to be known \chb{\em a priori}, ii) caching probability, which is dictated by the availability of the popular contents, and iii) coverage probability that guarantees  the minimum $\sir$ for successful reception.  
  \chm{Recalling} that \chm{the} caching probability  and the request probability for content $c_j$ were denoted by $b_j$ and ${\tt P}_{R_j}$,  the total hit probability can be mathematically expressed as:
 \begin{align}
{\tt P}_{\rm hit}=\sum_{j=1}^{{\cal J}}  {\tt P}_{R_j}  \sum_{k=1}^{N_{\rm t}} {\tt P}^{ (k)}_{{\rm c}} (1-b_j)^{k-1} b_j
\end{align}
where $(1-b_j)^{k-1} b_j $ indicates that the closest device with content of interest is $k^{th}$ closest device to the target-Rx. In other words,
    the content of interest was not found at $(k-1)$ closet transmitting devices to the target-Rx. \chm{On the same lines as \cite{BlaszczyszynG14}, the problem of optimal geographic caching in finite wireless networks can be formulated as:}
       \begin{align}
&{\tt P}_{\rm hit}^{*}=   \max_{\{b_j\}} \sum_{j=1}^{{\cal J}}  {\tt P}_{R_j}  \sum_{k=1}^{N_{\rm t}} {\tt P}^{(k)}_{{\rm c}} (1-b_j)^{k-1} b_j\\ \label{eq: constraint}
&\text{s.t.}\quad   \sum_{j=1}^{\cal J} b_j \le m_{\rm c}; \quad  0\le b_j \le 1\:\: \forall j
\end{align}
\chm{where $P_c^{(k)}$ is the new coverage probability result derived in Theorem \ref{thm: coverage k-closest}. The necessity and sufficiency of the constraints given by \eqref{eq: constraint} has already been discussed in \cite{BlaszczyszynG14}.}
 For better understanding of this optimization problem, \figref{Fig:optimal hit prob} plots the total hit probability for the simple setup of ${\cal J}=2$, and $m_{\rm c}=1$.  It can be seen that by increasing the number of simultaneously active transmitting devices, $N_{\rm a}$,
 the optimal caching probability for  the most popular content (i.e., $b_1$)  moves toward one. This is mainly because increasing $N_{\rm a}$ results in \chm{higher} interference, \chm{which in turn} decreases coverage probability.  For example when there is only one active device (\chb{completely} orthogonal channel allocation), coverage probability is equal to {\em one} under interference limited regime.  Hence, it is optimal to \chm{cache the two contents with the same probability, i.e.,} $b_1=b_2=0.5$. However, by increasing the number of active transmitting devices, ${\tt P}_{\rm c}^{(k)}$ for $k>1$ decreases significantly due to \chm{increased} interference. \chm{As demonstrated in \figref{Fig:optimal hit prob}, this makes it optimal to cache the more popular contents more often.}
 Though  channel orthogonalization is beneficial in terms of total hit probability, it is not \chb{desirable} for the network throughput which \chb{favors} having multiple active links as long as \chb{the} resulting interference remains acceptable. To study the tradeoff between \chb{the} number of active links and the resulting interference, we formally define throughput as: $ {\tt T}^{*}=N^{\rm a}\:{\tt P}_{\rm hit}^*
.$\\
% \begin{align}
%{\tt T}^{*}=N^{\rm a}\:{\tt P}_{\rm hit}^*
% \end{align}
Figs.~\ref{Fig:optimal hit prob vs number of active} and~\ref{Fig:optimal ASE vs number of active} plot maximum hit probability and  throughput versus number of active transmitting devices, respectively. Interestingly, increasing the  number of active devices has a conflicting effect on the maximum hit probability and the throughput: maximum hit probability decreases and throughput increases. %This shows that more and more  devices can be simultaneously active as long as the hit probability remains acceptable.
 This implies that more devices can be simultaneously activated if the \chg{total} hit probability remains within acceptable limits.

  \section{Conclusion}
Modeling the locations of the devices as a uniform-BPP,  \chm{we derive a new coverage result to enable the analysis of} optimal  \chb{geographic} caching in finite wireless networks. 
To the best of our knowledge, this is the first stochastic geometry-based analysis of  cache-enabled \chb{finite D2D networks}.
  This work has many extensions. From modeling perspective, it can be extended to an arbitrary shape \chb{(instead of a circle)} \chm{where the node locations follow a more general distribution.} Further, this framework can be  extended to analyze Mat\'ern cluster process where each finite network form one cluster. From system perspective, it can be used for the performance analysis of indoor communication and hotspots, where \chg{the analyses based on the infinite PPP assumption may not be applicable.}

\appendix
\subsection{Proof of Lemma \ref{lem: density function of interferer distance k}}
\label{App: i.i.d. property of u_in and u_out}
The joint density function of ``ordered" subset $\{W_{i:N^{\tt t}}\}_{i=1:N^{\tt t}}$ conditioned on $r=w_{k:N^{\tt t}}$  is:
\begin{align}\notag %\label{eq: joint con nu r}\
&f({{w_{1:N^{\tt t}}},...,{w_{N^{\tt t}:N^{\tt t}}}|w_{k:N^{\tt t}}})=\frac{f({{w_{1:N^{\tt t}}},...,{w_{N^{\tt t}:N^{\tt t}}}})}{f_{W_{k:N^{\tt t}}}(w_{k:N^{\tt t}})},\\\notag
&\stackrel{(a)}{=}\frac{N^{t}! \prod_{i=1}^{N^{t}}f_{W_i}(w_{i:N^{\tt t}})}{f_{W_{k:N^{\tt t}}}(w_{k:N^{\tt t}})}
\stackrel{(b)}{=}\underbrace{(k-1)! \prod_{i=1}^{k-1} \frac{f_{W_i}(w_{i:N^{\tt t}})}{F_W(w_{k:N})} }_{\text{joint PDF of } \{W_{i:N^{\tt t}}\}_{i=1:k-1} }\\\notag
& \times \underbrace{({N^{t}-k})! \prod_{i=k+1}^{N_{\tt t}} \frac{f_{W_i}(w_{i:N^{\tt t}})}{1-F_W(w_{k:N})}}_{\text{joint PDF of } \{W_{i:N^{\tt t}}\}_{i=k+1:N^{\tt t}} }
\end{align}
where $f_{W_{k:N^{\tt t}}}(.)=f^{(k)}_R(.)$.  Here $(a)$ follows from definition of the joint PDF of order of statistics \cite{ahsanullah2005order} with sampling PDF $f_{W_i}(.)$ given by \eqref{Eq: CDF unif}, and (b) follows from  substituting the expression of  $f_{W_{k:N^{\tt t}}}(.)$ which is given by \eqref{Eq: CDF unif}. The product of  joint PDFs in $(b)$   implies that the  $\{W_{i:N^{\tt t}}\}_{i=1:k}$  and $\{W_{i:N^{\tt t}}\}_{i=k+1:N^{\tt t}}$ conditioned on the  serving distance $r=w_{k:N}$  are independent. Now,  the joint PDF of $\{W_{i:N^{\tt t}}\}_{i=1:k}$  conditioned on $r=w_{k:N^{\tt t}}$  is:
\begin{align*}
f({{w_{1:N^{\tt t}}},...,{w_{k-1:N^{\tt t}}}|w_{k:N^{\tt t}}})=(k-1)! \prod_{i=1}^{k-1} \frac{f_{W_i}(w_{i:N^{\tt t}})}{F_{W_i}(w_{k:N^{\tt t}})}.
\end{align*}
Note that $(k-1)!$ shows possible permutations of distances from ``ordered'' subset, $\{w_{i:N^{\tt t}}\}_{i=1:k-1}$, and \chb{ doesn't appear }in the joint PDF of  distances from ``unordered'' subset $\ncalB^{\tt in}$, i.e.,  
%It's worth to note that it is not necessary to ``order'' the distances from interfering devices to target-Rx.
 % and hence \chb{does not   appear on} the joint PDF of  distances from ``unordered'' subset $\ncalB^{\tt in}$ that is: 
\begin{align*}
f({{w_{1}},...,{w_{k-1}}|w_{k:N^{\tt t}}})= \prod_{i=1}^{k-1} \frac{f_{W_i}(w_{i})}{F_{W_i}(w_{k:N^{\tt t}})},\:\: w_i\leq w_{k:N^{\tt t}}
\end{align*}
where the product of  same functional form of the joint PDF $f({{w_{1}},...,{w_{k-1}}|w_{k:N^{\tt t}}})$ infers that the subset of distances in ``unordered"  set $\ncalB_{\tt in}$ are i.i.d., with PDF $\frac{f_{W_i}(w_{i})}{F_{W_i}(w_{k:N^{\tt t}})}$. Denoting the distances in ``unordered" subset  $\ncalB^{\tt in}$ by $u_{\tt in}$, and substituting $w_{k:N^{\tt t}}$ with $r$ completes the proof.  Similar arguments can be applied for the derivation of  $f_{U_{\rm out}}(.|r)$.
\subsection{Proof of Lemma \ref{lemma: Laplace under k-closest}}
\label {App : Laplace k-closet}
The Laplace transform of interference conditioned on   $r=w_{k:N^{t}}$ is $\ncalL_{\ncalI}^{(k)}(s|r)$
\begin{align*}
%&=\E\Big[\exp\Big(-s\sum_{i=1,i\neq \ell}^{N^{\rm t}}  h_i\|\nbx_0+\nby_i\|^{-\alpha}\Big)\Big]\\
&=\E\Big[\prod_{u_{\tt in} \in \ncalB^{\tt in}} \exp\Big(-s  h_i {u_{\tt in}}^{-\alpha}\Big) \prod_{u_{\tt out}\in \ncalB^{\tt out}}\exp\Big(s  h_i {u_{\tt out}}^{-\alpha}\Big)\Big]\\
&\stackrel{(a)}=\E\Big[\prod_{u_{\tt in} \in \ncalB^{\tt in}} \frac{1}{1+s   {u_{\tt in}}^{-\alpha}} \prod_{u_{\tt out} \in \ncalB^{\tt out}}\frac{1}{1+s   {u_{\tt out}}^{-\alpha}}\Big]\\
&\stackrel{(b)}=\sum_{\ell=0}^{n^{\rm a}_{m}}\frac{p^\ell (1-p)^{N^{\rm a}-\ell-1} \binom{N^{\rm a}-1}{\ell}} {\sum_{\ell=0}^{N^{\rm a}_{m}}p^\ell (1-p)^{N^{\rm a}-\ell-1} \binom{N^{\rm a}-1}{\ell}} \Bigg(\int_0^{r} \frac{1}{1+s   {u_{\tt in}^{-\alpha} }}\times\\ & 
 f_{U_{\tt in}}(u_{\tt in}|r) \nrmd u_{\tt in} \Bigg)^{\ell} 
\Bigg(\int_r^{r_\nrmd}\frac{1}{1+s   {u_{\tt out}^{-\alpha}}}  f_{U_{\tt out}}(u_{\tt out}| r) \nrmd u_{\tt out} \Bigg)^{N^{\rm a}-\ell-1} 
\end{align*}
where $(a)$ follows from  $h_i \sim \exp(1)$, and $(b)$ follows from the fact that $U_{\tt in}\in \ncalB^{\tt in}$ and $U_{\tt out}\in \ncalB^{\tt out}$  are conditionally i.i.d., 
with PDF {\small $f_{U_{\tt in}}(.| r) $}, and  {\small $f_{U_{\rm out}}(.| r)$}, followed  by expectation over the number of devices located in ${\cal B}_{\tt in}$ and ${\cal B}_{\tt out}$, followed by the fact that the number of device in ${\cal B}_{\tt in}$ is binary random variable with probability $p=\frac{k-1}{N^{\tt t}-1}$ conditioned on total being less than $n^{\rm a}_{m}=\min(k-1,N^{\rm a}-1)$.  The integral can be solved based on \cite[{eq (3.194.1)}]{zwillinger2014table}.
%\chb{Substituting table of integral completes proof. }

%%%%%%%%%%%%%%%%%%%%%%
\bibliographystyle{IEEEtran}
\bibliography{ref,MA_pub}

\end{document}